\documentclass[fleqn,usenatbib,usedcolumn]{mnras}

\usepackage[british]{babel}             
\usepackage{newtxtext}                  
   \usepackage[slantedGreek]{newtxmath}    
   \usepackage[T1]{fontenc}                
   \usepackage{graphicx}                   
\renewcommand{\vec}[1]{\boldsymbol{#1}}

\title[Nested sampling on non-trivial geometries]{Nested sampling on non-trivial geometries}

\author[K. Javid]{
Kamran Javid$^{1,2}$\thanks{E-mail: kj316@mrao.cam.ac.uk}
\\
$^{1}$Astrophysics Group, Cavendish Laboratory, JJ Thomson Avenue, Cambridge CB3 0HE, UK\\
$^{2}$Kavli Institute for Cosmology, Madingley Road, Cambridge CB3 0HA, UK
}

\date{Accepted XXX. Received YYY; in original form ZZZ}

\pubyear{2018}

\begin{document}
\label{firstpage}
\pagerange{\pageref{firstpage}--\pageref{lastpage}}
\maketitle

\begin{abstract}
Metropolis nested sampling evolves a Markov chain from a current livepoint and accepts new points along the chain according to a version of the Metropolis acceptance ratio modified to satisfy the likelihood constraint, characteristic of nested sampling algorithms.
The geometric nested sampling algorithm we present here is a based on the Metropolis method, but treats parameters as though they represent points on certain geometric objects, namely circles, tori and spheres. For parameters which represent points on a circle or torus, the trial distribution is `wrapped' around the domain of the posterior distribution such that samples cannot be rejected automatically when evaluating the Metropolis ratio due to being outside the sampling domain. Furthermore, this enhances the mobility of the sampler. 
For parameters which represent coordinates on the surface of a sphere, the algorithm transforms the parameters into a Cartesian coordinate system before sampling which again makes sure no samples are automatically rejected, and provides a physically intutive way of the sampling the parameter space. \\
We apply the geometric nested sampler to two types of toy model which include circular, toroidal and spherical parameters. We find that the geometric nested sampler generally outperforms \textsc{MultiNest} in both cases. \\
Our implementation of the algorithm can be found at \url{https://github.com/SuperKam91/nested_sampling} \citep{javid2020geometric}.
\end{abstract}



\section{Introduction}
\label{s:gns_intro}
Bayesian inference is a vital tool for any astrophysicist who wants to obtain probabilistic estimates of model parameters from a dataset, whilst incorporating \textit{prior} information about the parameters (e.g. from another experiment). Bayesian inference can also be used to perform model comparisons by calculating values for the Bayesian \textit{evidence}, a quantity obtained by averaging the \textit{likelihood} function over the prior. The evidence incorporates Occam's razor: if two models fit the data equally well, the less complex model will be preferred. For most astrophysical problems, calculating the evidence numerically is infeasible, especially for high dimensional problems. Likewise, attempting to calculate parameter probability distributions exactly is computationally impossible. Thus one usually resorts to statistical sampling to gain estimates of these quantities. \\
\citet{2004AIPC..735..395S} introduced a novel sampling method referred to as nested sampling. This algorithm focuses on calculating the evidence, but also generates samples from the \textit{posterior} probability distribution. The key computational expense associated with nested sampling is the constraint that newly generated samples must be above a certain likelihood value which increases at each iteration. Nested sampling has proved to be a very popular choice for Bayesian inference in the astrophysics community, due to its ability to calculate Bayesian evidence values for model comparison as well as produce posterior distributions (see e.g. \citealt{2018arXiv180501968J}, \citealt{2019MNRAS.483.3529J}, \citealt{2019MNRAS.486.2116P}, and \citealt{2019MNRAS.489.3135J} which all use data from the Arcminute MicroKelvin Imager \citealt{2018MNRAS.475.5677H} to perform model comparison and posterior analysis using nested sampling). \\
Initially, \citet{Sivia2006} suggested satisfying this constraint by evolving a Markov chain starting at one of the pre-existing samples and evaluating an acceptance ratio based on the one used by the Metropolis algorithm \citep{1953JChPh..21.1087M} used in Markov Chain Monte Carlo (MCMC) sampling (see e.g. \citealt{mackay2002} for a review). A variant of the nested sampling algorithm which focused on sampling from ellipsoids which approximate the region in which the likelihood constraint is satisfied was also developed \citep{2006ApJ...638L..51M}. A major breakthrough in the applicability of nested sampling to highly multi-modal distributions came with the invention of clustering nested sampling algorithms (\citealt{2007MNRAS.378.1365S}, \citealt{2008MNRAS.384..449F}, and \citealt{2009MNRAS.398.1601F}). These algorithms effectively sample from multiple ellipsoids determined by some clustering algorithm, with the aim of approximating the likelihood constraint for each mode of the distribution.
More recently, the slice sampling algorithm \textsc{POLYCHORD} (\citealt{2015MNRAS.450L..61H}, \citealt{2015MNRAS.453.4384H}) has been introduced and is effective at navigating high dimensional spaces, due to the fact that it is not a rejection sampling algorithm. Section~4.1 of \citet{2015MNRAS.453.4384H} gives further examples of nested sampling algorithms which have different ways of satisfying the likelihood constraint.

Here we introduce a nested sampling algorithm which introduces a new variant for satisfying the likelihood constraint, based on the Markov method used in \citet{Sivia2006} (and also applied in \citealt{2008MNRAS.384..449F}). Certain parameters relevant to astrophysics exhibit special properties which mean they naturally parameterise points on geometric objects such as circles, tori and spheres. The algorithm we introduce here which we refer to as the geometric nested sampler, exploits these properties to generate samples efficiently and enables mobile exploration of distributions which are defined on such geometries. 

In Sections~\ref{s:gns_bayes} and~\ref{s:gns_ns} we give more formal introductions of Bayesian inference and nested sampling respectively. In Section~\ref{s:gns_lhood_const} we give a more detailed account on the Markov chain method used in \citet{Sivia2006} and \citet{2008MNRAS.384..449F} to satisfy the likelihood constraint, before introducing the variation of this method used in geometric nested sampling in Section~\ref{s:gns_geom_ns}. We then apply the geometric nested sampling algorithm to circular, toroidal and spherical toy models in Sections~\ref{s:gns_tmI} and~\ref{s:gns_tmII}. 
Finally, we provide conclusions of our work in Section~\ref{s:gns_conc}.

\section{Bayesian inference}
\label{s:gns_bayes}

For a model $\mathcal{M}$ and data $\vec{\mathcal{D}}$, we can obtain model parameters (also known as input or sampling parameters) $\vec{\theta}$ conditioned on $\mathcal{M}$ and $\vec{\mathcal{D}}$ using Bayes' theorem:
\begin{equation}\label{e:gns_bayes}
\mathrm{Pr}\left(\vec{\theta}|\vec{\mathcal{D}},\mathcal{M}\right) = \frac{\mathrm{Pr}\left(\vec{\mathcal{D}}|\vec{\theta},\mathcal{M}\right)\mathrm{Pr}\left(\vec{\theta}|\mathcal{M}\right)}{\mathrm{Pr}\left(\vec{\mathcal{D}}|\mathcal{M}\right)},
\end{equation}
where $\mathrm{Pr}\left(\vec{\theta}|\vec{\mathcal{D}},\mathcal{M}\right) \equiv \mathcal{P}\left(\vec{\theta}\right)$ is the posterior distribution of the model parameter set, $\mathrm{Pr}\left(\vec{\mathcal{D}}|\vec{\theta},\mathcal{M}\right) \equiv \mathcal{L}\left(\vec{\theta}\right)$ is the likelihood function for the data, $\mathrm{Pr}\left(\vec{\theta}|\mathcal{M}\right) \equiv \pi\left(\vec{\theta}\right)$ is the prior probability distribution for the model parameter set, and $\mathrm{Pr}\left(\vec{\mathcal{D}}|\mathcal{M}\right) \equiv \mathcal{Z}$ is the Bayesian evidence of the data given a model $\mathcal{M}$. The evidence can be interpreted as the factor required to normalise the posterior over the model parameter space:
\begin{equation}\label{e:gns_evidence}
\mathcal{Z} = \int \mathcal{L}\left(\vec{\theta}\right) \pi\left(\vec{\theta}\right)\, \mathrm{d}\vec{\theta}.
\end{equation} 
$\mathcal{Z}$ is central to the way in which nested sampling algorithms determine samples from $\mathcal{P}\left(\vec{\theta}\right)$, and so they are often used to calculate accurate estimates of both $\mathcal{Z}$ and $\mathcal{P}\left(\vec{\theta}\right)$.


\subsection{Prior distributions}
\label{s:gns_priors}
In general for nested sampling $\pi\left(\vec{\theta}\right)$ can take any form as long as the distribution integrates to one and has a connected support (\citealt{2008arXiv0801.3887C}, informally this means that the parts of the domain at which $\pi\left(\vec{\theta}\right) \neq 0$ is not `separated' by the parts at which $\pi\left(\vec{\theta}\right) = 0$). For simplicity, in all examples considered in this paper (except for parameters which describe the azimuthal angle of a sphere, see below) we assume that each component of the $N$-dimensional vector $\vec{\theta}$ is independent of one another and that each $\pi(\theta_i)$ is a uniform probability distribution so that
\begin{equation}\label{e:gns_priors}
\pi\left(\vec{\theta}\right) = \prod_{i=1}^{N} \pi_{i}(\theta_{i}) = \prod_{i=1}^{N} \frac{1}{\theta_{\mathrm{max},i} - \theta_{\mathrm{min},i}},
\end{equation}
where $\theta_{\mathrm{max},i}$ and $\theta_{\mathrm{min},i}$ are respectively the upper and lower bounds on $\theta_i$. Values for $\theta_{\mathrm{max},i}$ and $\theta_{\mathrm{min},i}$ used in the examples presented here will be stated in the relevant Sections where the examples are introduced (i.e. Sections~\ref{s:gns_tmI} and~\ref{s:gns_tmII}).
In the case of parameters which resemble azimuthal angles, it is appropriate to assign a sinusoidal prior,
\begin{equation}\label{e:sin_prior}
\pi_{i}(\theta_{i}) = \frac{1}{\theta_{\mathrm{max},i} - \theta_{\mathrm{min},i}} \sin \left(\frac{\pi \theta_i}{\theta_{\mathrm{max},i} - \theta_{\mathrm{min},i}} \right).
\end{equation}
This ensures that for spherical distributions (see Section~\ref{s:gns_tmII}), the prior distribution integrates to one over the surface of a sphere.


\section{Nested sampling} 

\label{s:gns_ns}
\begin{figure}
  \begin{center}
  \includegraphics[ width=0.90\linewidth]{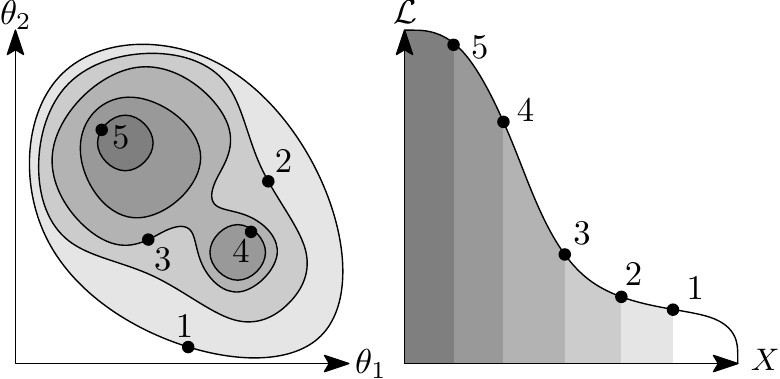}
  \caption{Left plot: Five iso-likelihood contours of a two-dimensional, multi-modal likelihood $\mathcal{L}(\theta_1, \theta_2)$. Each contour encloses some fraction of the prior $X$, with the colourscale indicating the value of $X$ (darkest: smallest $X$). Right: Corresponding $\mathcal{L}$ as a function of $X$ plot (not to scale). The area under the curve is equal to $\mathcal{Z}$.} 
  \label{f:gns_ns_plots}
  \end{center}
\end{figure}

Nested sampling exploits the relation between the likelihood and `prior volume' to transform the multidimensional integral given by equation~\ref{e:gns_evidence} into a one-dimensional integral. The prior volume $X$ is defined by $\mathrm{d}X = \pi\left(\vec{\theta}\right) \mathrm{d}\vec{\theta}$, thus $X$ is defined on $[0,1]$ and we can set
\begin{equation}
\label{e:gns_ns}
X(\mathcal{L}) = \int_{\mathcal{L}\left(\vec{\theta}\right) > \mathcal{L}} \pi\left(\vec{\theta}\right) \mathrm{d}\vec{\theta}.
\end{equation}
The integral extends over the region(s) of the parameter space contained within the iso-likelihood contour $\mathcal{L}\left(\vec{\theta}\right) = \mathcal{L}$ (see Figure~\ref{f:gns_ns_plots}). Assuming that the inverse of equation~\ref{e:gns_ns} ($\mathcal{L}(X) \equiv X^{-1}(\mathcal{L})$) exists which is the case when $\pi$ is strictly positive, then the evidence integral can be written as (a proof of this equivalence is given in Appendix~D of \citealt{kj_thesis})
\begin{equation}\label{e:gns_nsz}
\mathcal{Z} = \int_{0}^{1} \mathcal{L}(X) \mathrm{d}X.
\end{equation}
Thus, if one can evaluate $\mathcal{L}(X)$ at $n_{\rm s}$ values of $X$, the integral given by equation~\ref{e:gns_nsz} can be approximated by standard quadrature methods
\begin{equation}\label{e:gns_nsz_sum}
\mathcal{Z} \approx \sum_{i = 1}^{n_{\rm s}} \mathcal{L}_{i} (X_{i-1} - X_{i}),
\end{equation}
where 
\begin{equation}\label{e:gns_X_vals}
0 < X_{n_{\rm s}} < ... < X_{1} < X_{0} = 1.
\end{equation}

\subsection{Distribution of $\mathcal{Z}$} 
As explained in \citet{2004AIPC..735..395S}, the geometric uncertainty associated with the $X_i$ leads to the idea that $\log(\mathcal{Z})$ rather than $\mathcal{Z}$ is a normally distributed variable. Assuming the latter to be normally distributed can result in distributions of $\mathcal{Z}$ with variances that suggest $\mathcal{Z}$ can take negative values, which is unphysical. This is the case with the spherical toy model considered in Section~\ref{s:gns_tmII}.
The mean and variance of a log-normally distributed random variable, $\mathbb{E}\left[ \log(\mathcal{Z}) \right]$ and $\mathrm{var}\left[ \log(\mathcal{Z}) \right]$, can be calculated from the moments of the non-logarithmic variables as
\begin{gather} 
\label{e:gns_lognorm_mean}
\mathbb{E}\left[ \log(\mathcal{Z}) \right] = 2 \log \left( \mathbb{E}[\mathcal{Z}] \right) - \frac{1}{2} \log \left( \mathbb{E}\left[\mathcal{Z}^2\right] \right), \\
\label{e:gns_lognorm_var}
\mathrm{var}\left[ \log(\mathcal{Z}) \right] = \log \left( \mathbb{E}\left[ \mathcal{Z}^2 \right] \right) - 2 \log \left( \mathbb{E}[\mathcal{Z}] \right).
\end{gather}
Hence our geometric nested sampling algorithm calculates the moments of the linear variables
, but the final evidence estimate and its associated error are calculated using equations~\ref{e:gns_lognorm_mean} and~\ref{e:gns_lognorm_var}.  

\subsection{Stopping criterion}
\label{s:gns_stop_crit}
The nested sampling algorithm can be terminated based on an estimate of how precisely the evidence value has been calculated up to the current iteration. One measure of this is to look at the ratio of the current estimate of $\mathcal{Z}$ to its value plus an estimate of the `remaining' evidence associated with the current livepoints. Since after iteration $n_{\rm s}$ the livepoints are uniformly distributed in the range $[0, X_{n_{\rm s}}]$, we can approximate their final contribution to the evidence as
\begin{equation}\label{e:ngs_final}
\mathcal{Z}_{\rm f} \approx \frac{X_{n_{\rm s}}}{n_{l}} \sum_{i = 1}^{n_{l}} \mathcal{L}_i,
\end{equation}
where $\mathcal{L}_i$ is the likelihood value of the $i^{\mathrm{th}}$ remaining livepoint. The stopping criterion can then be quantified as
\begin{equation}\label{e:ngs_stop_crit}
\frac{\mathcal{Z}_{\rm f}}{\mathcal{Z}_{\rm f} + \mathcal{Z}} < \epsilon. 
\end{equation} 
$\epsilon$ is a user defined parameter, which we set to $0.01$ in our implementation.

\subsection{Posterior inferences}
\label{s:gns_post_points}
Once $\mathcal{Z}$ has been determined, posterior inferences can easily be generated using the deadpoints and final livepoints from the nested sampling process to give a total of $n_{\rm s} + n_{l}$ samples. Each such point is assigned the weight
\begin{equation}\label{e:ngs_post_weights}
\mathcal{P}_{i} = \frac{\mathcal{L}_{i} \left(X_{i-1} - X_{i}\right)}{\mathcal{Z}}.
\end{equation}
The weights (along with the corresponding values of $\vec{\theta}$) can be used to calculate statistics of the posterior distribution, or plot it using software such as \textsc{getdist}\footnote{\url{http://getdist.readthedocs.io/en/latest/}.}.


\section{Satisfying the likelihood constraint} 
\label{s:gns_lhood_const}
At each step of the nested sampling iteration, one needs to sample a new point which satisfies $\mathcal{L}_{\rm t} > \mathcal{L}_{i}$. As mentioned in the Introduction, considerable work has been put into increasing the efficiency of this process, as it is by far the most computationally expensive step of the nested sampling algorithm. We now give a review of the Metropolis likelihood sampling methodology used by \citet{Sivia2006} and \citet{2008MNRAS.384..449F}, which forms the basis of the method used in geometric nested sampling.

\subsection{Metropolis likelihood sampling}
\label{s:gns_m_lhood_const}
The Metropolis nested sampling method is an adaption of the algorithm used in MCMC sampling of a posterior distribution \citep{1953JChPh..21.1087M}. The acceptance ratio $\alpha$ for the Metropolis algorithm can be derived from the \textit{detailed balance} relation
\begin{equation}
\label{e:gns_db}
\mathcal{P}\left(\vec{\theta}_{j}\right) T\left(\vec{\theta}_{j+1} | \vec{\theta}_{j}\right) = \mathcal{P}\left(\vec{\theta}_{j+1}\right) T\left(\vec{\theta}_{j} | \vec{\theta}_{j+1}\right),
\end{equation}
where $T\left(\vec{\theta}_{j+1} | \vec{\theta}_{j}\right)$ denotes the probability of transitioning from state $\vec{\theta}_{j}$ to state $\vec{\theta}_{j+1}$ in one step along a Markov chain. A derivation of $\alpha$ for the standard Metropolis algorithm and more information on the transition distribution are given in \citet{kj_thesis}. For more information on Markov processes in general we refer the reader to \citet{mackay2002} and \citet{robert_casella2004}.
The acceptance ratio for the Metropolis nested sampling algorithm takes the form
\begin{equation}
\label{e:gns_ns_accept}
\alpha = \begin{cases}
\mathrm{min}\left[\pi\left(\vec{\theta}_{\rm t}\right) / \pi\left(\vec{\theta}_{l}\right), 1\right] \quad \mathrm{ if } \quad \mathcal{L}_{\rm t} > \mathcal{L}_{i}, \\
 0 \quad \mathrm{ otherwise.}
\end{cases}
\end{equation}
Here $\vec{\theta}_{l}$ is obtained by picking one of the current livepoints at random, and using its value of $\vec{\theta}$. The value for $\vec{\theta}_{\rm t}$ is sampled from a trial distribution $q\left(\vec{\theta}_{\rm t} | \vec{\theta}_{l}\right)$. Sivia \& Skilling and Feroz et al. use symmetric Gaussian distributions centred on $\vec{\theta}_{l}$ for $q\left(\vec{\theta}_{\rm t} | \vec{\theta}_{l}\right)$. The trial point is accepted to be a new livepoint (replacing the deadpoint associated with $\mathcal{L}_{i}$) with probability $\alpha$. Note that equation~\ref{e:gns_ns_accept} implicitly assumes that the proposal distribution is symmetric in its arguments, that is $q\left(\vec{\theta}_{\rm t} | \vec{\theta}_{l}\right) = q\left(\vec{\theta}_{l} | \vec{\theta}_{\rm t}\right)$. In the case that the proposal distribution is asymmetric, the acceptance ratio includes an additional factor $q\left(\vec{\theta}_{l} | \vec{\theta}_{\rm t}\right)q\left(\vec{\theta}_{\rm t} | \vec{\theta}_{l}\right)$ (in which case the algorithm is referred to as the Metropolis-Hastings algorithm, \citealt{1970Bimka..57...97H}).

The fact that the Metropolis nested sampling method uses the current livepoints as a `starting point' for selecting $\vec{\theta}_{\rm t}$, means that the autocorrelation between the livepoints is high, which in turn leads to biased sampling. This can be prevented by increasing the variance of the trial distribution used, or by requiring that multiple trial points must be accepted before the final one is accepted as a livepoint, i.e. after the first accepted trial point is found, set $\vec{\theta}_{l} \rightarrow \vec{\theta}_{\rm t}$ and use this to sample a new $\vec{\theta}_{\rm t}$ from $q\left(\vec{\theta}_{\rm t} | \vec{\theta}_{l}\right)$. This can be repeated an arbitrary number of times, but in general more iterations leads to a lower correlation between the livepoint used at the beginning of the chain and the final accepted trial point which is added to the livepoint set. \\ Sivia \& Skilling suggest that at each nested sampling iteration, the number of trial points generated $n_{\rm t}$ to get a new livepoint should be $\approx 20$. In our implementation we set this number to $20 \times N$ where $N$ is the dimensionality of the parameter estimation problem. Note that $n_{\rm t}$ includes both accepted and rejected trial points. Sivia and Skilling also suggest that the acceptance rate for the trial points at each nested sampling iteration should be $\approx 50\%$. This is because a high acceptance rate usually suggests high auto-correlation between the successive trial points, whilst a low acceptance rate can suggest high correlation between the final accepted trial point and the one used to initialise the chain, as too few steps have been made between the two. In the extreme case that the acceptance rate is zero, the process of picking a new livepoint has failed, as one cannot have two livepoints corresponding to the same $\vec{\theta}$. The acceptance rate is affected by the variance of the trial distribution, a large variance usually results in more trial points being rejected (especially near the peaks of the posterior). Sivia \& Skilling suggest updating the trial standard deviation as
\begin{equation}
\label{e:gns_trialsig}
\sigma_{\rm t} \rightarrow \begin{cases}
 \sigma_{\rm t} \exp(1/N_{\rm a}) \quad \mathrm{ if } \quad N_{\rm a} > N_{\rm r}, \\
 \sigma_{\rm t} \exp(-1/N_{\rm r}) \quad \mathrm{ if } \quad N_{\rm a} \leq N_{\rm r},
\end{cases}
\end{equation}
where $N_{\rm a}$ and $N_{\rm r}$ are the number of accepted and rejected trial points in the current nested sampling iteration respectively. Note however that we determine the variance using different methods (see Sections~\ref{s:gns_trialvar} and~\ref{s:gns_spherevar}).

Feroz et al. incorporate the Metropolis likelihood sampling into their clustering nested sampling algorithm rather than use it in isolation. The geometric likelihood sampling we introduce in the next Section is a modified version of the Metropolis algorithm used in isolation. 

\section{Geometric nested sampling}
\label{s:gns_geom_ns}

\begin{figure}
  \begin{center}
  \includegraphics[ width=0.90\linewidth]{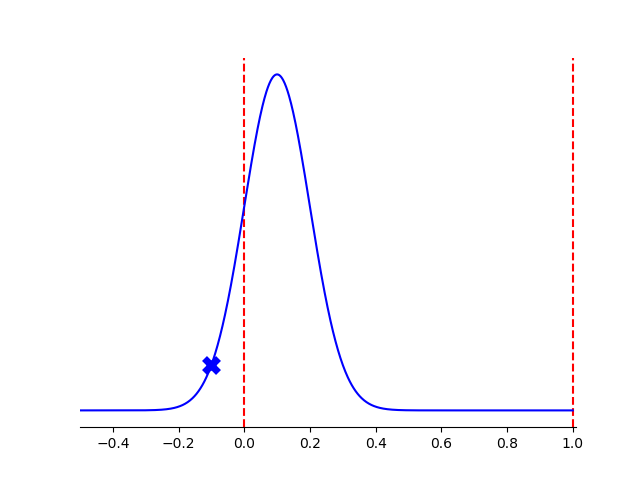}
  \caption{`Vanilla' non-wrapped trial distribution. The blue curve represents a Gaussian `vanilla' trial distribution $q(\theta'|\theta)$ with starting point $\theta = 0.1$, and sampled trial point $\theta' = -0.1$ shown by the blue cross.
	The support of $\pi$ is indicated by the red dashed lines ($[0,1]$). Since $\theta'$ lies outside the support of $\pi$,
	it would automatically be rejected by the Metropolis algorithm.} 
  \label{f:gns_nonwrap}
  \end{center}
\end{figure}

One key issue with Metropolis nested sampling is that at each nested sampling iteration, if too many trial points are rejected, then the livepoints will be highly correlated with each other after a number of nested sampling iterations. To prevent this one must sample a large number of trial points in order to increase the number of acceptances and decrease the auto-correlation of the trial point chain. \\ This solution can be problematic if computing the likelihood is computationally expensive. One particular case in which the sampled point is guaranteed to be rejected, is if the point lies outside of the domain of $\mathcal{P}$ (support of $\pi$). Such a case is illustrated in Figure~\ref{f:gns_nonwrap}. Of course, this can be avoided by adapting $q\left(\vec{\theta}_{\rm t} | \vec{\theta}_{l}\right)$ so that it is truncated to fit the support of $\pi$, but in high dimensions this can be tedious, and inefficient in itself. Hence one desires an algorithm which does not sample outside the support of $\pi$, without having to truncate $q$.

Another issue which most sampling algorithms are subject to occurs when the modes of the posterior distribution are far away from each other in $\vec{\theta}$ space, e.g. when they are at `opposite ends' of the domain of $\pi$. In the context of nested sampling this can result in one or more of the modes not being sampled accurately, particularly in the case of low livepoint runs.
Thus a sampling algorithm should be able to efficiently manoeuvre between well separated modes which lie at the `edges' of $\pi$'s support. 

Geometric nested sampling attempts to solve these two issues by interpreting parameter values as points on geometric objects, namely on circles, tori and spheres.

\subsection{Wrapping the trial distribution}
\label{s:gns_wrap}
\begin{figure}
  \begin{center}
  \includegraphics[ width=0.90\linewidth]{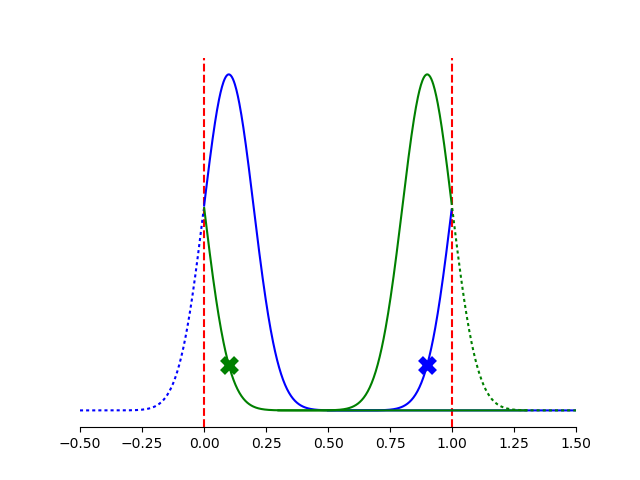}
  \caption{Wrapped trial distributions. The solid blue curve represents a Gaussian trial distribution $q(\theta'|\theta)$ as in previous Figure, but now incorporating the wrapping methodology. As a result of the wrapping, $\theta'$ (blue cross)
	is at $0.9$, and so won't be automatically rejected by the Metropolis algorithm.
	The green curve shows the same trial distribution $q(\theta|\theta')$ centred on $0.9$.
	The fact that $\theta =0.1$ (green cross) is sampled from $q(\theta|\theta')$ with the same probability as $\theta'$ is from $q(\theta'|\theta)$ shows that the wrapped trial distribution is still symmetric with respect to its arguments (provided $q(a|b)$ is a symmetric function about the point $b$).} 
  \label{f:gns_wrap}
  \end{center}
\end{figure}

A relatively straightforward way of ensuring that the trial points sampled from $q$ are in the support of $\pi$ is to `wrap' $q$. This is illustrated in Figure~\ref{f:gns_wrap}, where we consider a one-dimensional uniform prior on $[0, 1]$ for simplicity. 
For any point $\theta$, there will be a non-zero probability of sampling a value of $\theta'$ from the trial distribution $q(\theta' | \theta)$ that lies outside $[0, 1]$. If the point sampled has a value of say $\theta' = -0.1$, then if we consider $q$ to be wrapped around the support this can be interpreted as sampling a point at value $\theta' = 0.9$.
More generally, if $\theta'$ is outside the support of $\pi$ defined by upper and lower bounds $\theta_{\mathrm{max}}$ and $\theta_{\mathrm{min}}$ it will be transformed as
\begin{equation}
\label{e:gns_wrap}
\theta' = \begin{cases}
\theta_{\mathrm{max}} - W(\theta') \quad \mathrm{ if } \quad \theta' > \theta_{\mathrm{max}}, \\
\theta_{\mathrm{min}} + W(\theta') \quad \mathrm{ if } \quad \theta' < \theta_{\mathrm{min}},
\end{cases} 
\end{equation}
where
\begin{equation}
\label{e:gns_wrapmod}
W(\theta) = \begin{cases}
(\theta - \theta_{\mathrm{max}}) \mod (\theta_{\mathrm{max}} - \theta_{\mathrm{min}}) \quad \mathrm{ if } \quad \theta > \theta_{\mathrm{max}}, \\
(\theta_{\mathrm{min}} - \theta) \mod (\theta_{\mathrm{max}} - \theta_{\mathrm{min}}) \quad \mathrm{ if } \quad \theta < \theta_{\mathrm{min}}.
\end{cases}
\end{equation}
Assuming the support of $\pi$ is connected (a requirement of nested sampling, as stated in Section~\ref{s:gns_priors}), then this operation will be well defined for all $\pi$ with bounded supports, of arbitrary dimension. Using this transformation does not affect the argument symmetry of $q$, thus the value of $\alpha$ given by equation~\ref{e:gns_ns_accept} still holds. Furthermore, this symmetry ensures that the detailed balance relation given by equation~\ref{e:gns_db} is still satisfied.

\subsection{Circular parameters}
\label{s:gns_circular}

As well as ensuring that none of the sampled trial points lie outside the support of $\pi$, the wrapped trial distribution can also improve the manoeuvrability of the sampling process, since the trial point chain can always `move in either direction' without stepping outside of the support of $\pi$. This proves to be particularly useful for `circular parameters'. Here we define circular parameters to be those whose value at $\theta_{\mathrm{max}}$ and $\theta_{\mathrm{min}}$ correspond physically to the same point. Examples of circular parameters include angles (which are circular at e.g. zero and $2\pi$) and time periods (e.g. $00$:$00$ and $24$:$00$). 

\begin{figure}
  \begin{center}
  \includegraphics[ width=0.90\linewidth]{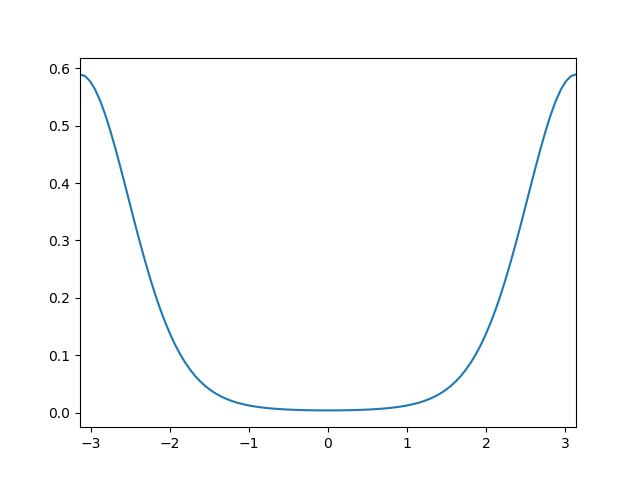}
  \caption{von Mises distribution with domain $[-\pi, \pi]$, centred on $\pi$. The peak wraps around at edges of domain, so that it appears as two half peaks on a linear space.} 
  \label{f:gns_vm}
  \end{center}
\end{figure}

Often, circular parameters have probability distributions associated with them which are also circular. An example of a circular distribution is the von Mises distribution, an example of which is shown in Figure~\ref{f:gns_vm} (and defined in Section~\ref{s:gns_tmI}). This particular example shows that the function's peak(s) may be split by the wrapping, so that when plotted linearly, they appear to have to `half peaks' about $\theta_{\mathrm{max}}$ and $\theta_{\mathrm{min}}$.
Such half peaks would be classified as two separate peaks by clustering nested sampling algorithms. Thus in general, the number of livepoints would need to be increased to accommodate for the higher number of modes, to ensure both half peaks are sampled adequately without one cluster `dying out'. Furthermore, the two half peaks occur at opposite ends of the domain of a linear space, making it more difficult for a sampler to explore the regions of higher probability efficiently. \\
The wrapped trial distribution resolves both of these issues, as the two half peaks in linear space are treated as one full peak as far as the sampling (and allocation of livepoints) is concerned. Consequently, the second issue of the half peaks being far away from each other is automatically eradicated. \\
The wrapped trial distribution methodology can thus be applied to problems which involve sampling on non-Euclidean spaces. We apply the method to toy models with distributions defined on circles and tori in Section~\ref{s:gns_tmI}.

\subsection{Variance of the trial distribution}
\label{s:gns_trialvar}
As with any sampling procedure which relies on a trial distribution, picking a variance for the distribution is difficult without a-priori knowledge of the posterior distribution you are sampling from. A low variance results in a lot of trial points being accepted, but a high auto correlation between these points. A high variance gives a lot of trial rejections, but when these points are accepted, their correlation with the starting point is often low.
Since picking the trial variance can in itself be a mammoth task, we use a simplistic approach and take it to be 
\begin{equation}
\label{e:gns_trialvar}
0.1 \times \left\lvert \max\displaylimits_{\mathrm{livepoints}}\left(\theta_{i}\right) - \min\displaylimits_{\mathrm{livepoints}}\left(\theta_{i}\right) \right\rvert ,
\end{equation}
for each component $i$ of $\vec{\theta}$. We use this approach to avoid the sampler from taking large steps when the livepoints are close together. However, we acknowledge that this method is far from optimal when the livepoints are compactly located at the edges of the domain of $\mathcal{P}\left(\vec{\theta}\right)$. 

\subsection{Non-Euclidean sampling via coordinate transformations}
\label{s:gns_coordtrans}
The wrapped trial distribution introduced in Section~\ref{s:gns_wrap} can in theory be used in Metropolis nested sampling to sample effectively from circular and toroidal spaces parameterised in terms of circular variables. However, it is not particularly effective at sampling from spherical spaces, since wrapping around the zenith angle (usually defined on $[0, \pi]$) would result in discontinuous jumps between the poles of the sphere. One could of course just wrap the trial distribution in the dimension representing the azimuthal angle (usually defined on $[0, 2\pi]$), rather than in both angles. However, this would re-introduce the issues stated in Section~\ref{s:gns_geom_ns}, i.e. wasting samples and inefficient exploration of the parameter space.
We therefore propose an alternative method for exploring spherical spaces which we incorporate in the geometric nested sampling algorithm.

\subsection{Spherical coordinate transformations}
\label{s:gns_spheretrans}
\begin{figure*}
  \begin{center}
  \includegraphics[ width=0.6\linewidth, keepaspectratio]{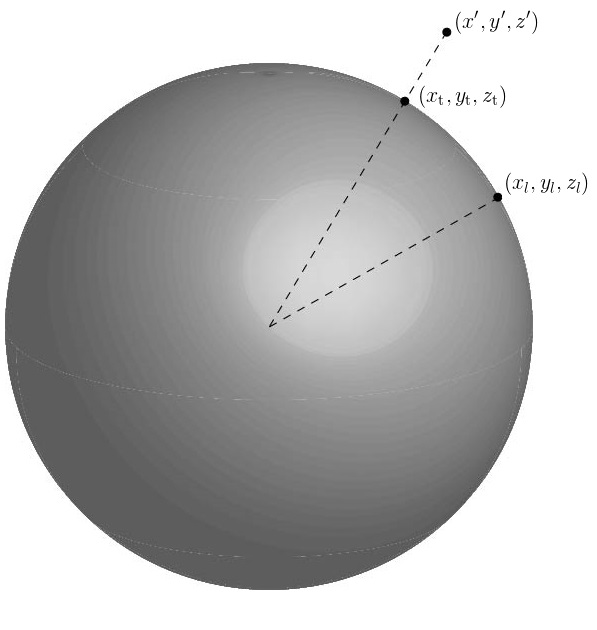}
  \caption{Sampling points on the surface of a sphere in Cartesian coordinates. The three-dimensional trial distribution is centred at the point $(x_l,y_l,z_l)$, which corresponds to $(\phi_l,\theta_l)$. The point $(x',y',z')$ sampled from $q$ in general will not lie on the surface of the sphere, however the point is implicitly projected onto the sphere at $(x_{\rm t},y_{\rm t},z_{\rm t})$ when calculating $(\phi',\theta')$ [ $\equiv (\phi_{\rm t},\theta_{\rm t})$].} 
  \label{f:gns_sphere}
  \end{center}
\end{figure*}

Assuming the surface of a unit sphere is parameterised by azimuthal angle $\phi$ on $[0, 2\pi]$ and zenith angle $\theta$ on $[0, \pi]$, then the corresponding Cartesian coordinates are
\begin{equation}
\label{e:gns_spherecoords}
\begin{split}
&x = r \cos(\phi)\sin(\theta), \\
&y = r \sin(\phi)\sin(\theta), \\
&z = r \cos(\theta),
\end{split}
\end{equation}
with $r = 1$. Note that $\phi$ is the angle measured anti-clockwise from the positive $x$-axis in the $x$--$y$ plane and $\theta$ is the angle measured from the positive $z$-axis. Thus a trial point $\phi_{\rm t}, \theta_{\rm t}$ can be sampled as follows. 
Starting from a point $\phi_{l}, \theta_l$, calculate $x_l, y_l, z_l$, from which a trial point $x', y', z'$ can be sampled from $q(x', y', z' | x_l, y_l, z_l)$. We use a three-dimensional spherically symmetric Gaussian distribution for $q(x', y', z' | x_l, y_l, z_l)$. In general, the point $x', y', z'$ will not lie on the unit sphere. Nevertheless the point is implicitly projected onto it by solving the equations given by~\ref{e:gns_spherecoords} simultaneously for $\phi$ and $\theta$, where we set $x = x'$, $y = y'$, $z = z'$, and $r = r'$ (see Figure~\ref{f:gns_sphere}). 
The resulting values are $\phi_{\rm t}$ and $\theta_{\rm t}$, from which the acceptance ratio given by equation~\ref{e:gns_ns_accept} can be evaluated as normal. \\ 
There are a few things to note about sampling the trial point in the Cartesian space. Firstly, for equation~\ref{e:gns_ns_accept} to hold we must have $q(\phi_{\rm t}, \theta_{\rm t} | \phi_l, \theta_l) = q(\phi_l, \theta_l | \phi_{\rm t}, \theta_{\rm t})$, which is equivalent to 
\begin{equation}
\label{e:gns_carttrials}
\int \displaylimits_{\vec{x}' \in \{\vec{x}_{\mathrm{t}, \phi,\theta}\}} q(\vec{x}'|\vec{x}) \mathrm{d}\vec{x}' = 
\int\displaylimits_{\vec{x} \in \{\vec{x}_{l,\phi,\theta}\}} q(\vec{x}|\vec{x}') \mathrm{d}\vec{x},
\end{equation}
where $\vec{x}' = (x',y',z')$ and $\vec{x} = (x,y,z)$. $\{\vec{x}_{\mathrm{t}, \phi,\theta}\}$ are the set of Cartesian coordinates which satisfy~\ref{e:gns_spherecoords} for $\phi = \phi_{\rm t}$, $\theta = \theta_{\rm t}$, and all $r \neq 0$. Similarly $\{\vec{x}_{l,\phi,\theta}\}$ are the $\vec{x}$ which satisfy~\ref{e:gns_spherecoords} for $\phi = \phi_l$ \& $\theta = \theta_l$ (see Figure~\ref{f:gns_sphere}). 
Due to the symmetry of the spherical coordinate system, these sets of vectors lie along the lines given by $(\phi_{\rm t}, \theta_{\rm t})$ and $(\phi_l, \theta_l)$ respectively. The only additional requirement for equation~\ref{e:gns_carttrials} to hold is that $q(x',y',z'|x,y,z)$ is symmetric in its arguments, which it is provided that $q(a|b)$ is a symmetric function about the point $b$.
As in Section~\ref{s:gns_wrap}, the symmetry of the trial distribution ensures that the detailed balance relation given by equation~\ref{e:gns_db} is still satisfied.

Sampling in Cartesian coordinates eliminates the risk of sampling points which are automatically rejected (due to being outside the support of $\pi(\phi,\theta)$) to a negligible level, since the only points in Cartesian coordinates which are ill-defined in spherical coordinates are $x=y=0$ for all $z$.
How the coordinate transformation improves the manoeuvrability of the sampler relative to sampling in the original parameter space is less clear-cut. For the latter, when the variance is fixed the step sizes taken by the sampler along the surface of the sphere depend on where you start from. For example, at $\theta \approx 0$, large moves in $\phi$ will result in relatively small steps along the sphere whereas at $\theta \approx \pi/2$ such moves in $\phi$ would result in large steps along the sphere. However when sampling in a Cartesian coordinate system, for a constant variance (see below), the trial points sampled will have the same average step size in Euclidean space regardless of the starting point. Furthermore due to the symmetry of a sphere, when the sampled point ($x',y',z'$) is projected back onto the sphere (implicitly when determining $\phi_{\rm t}$ and $\theta_{\rm t}$), the variance of the steps along the sphere is still independent of the starting point. \\
In either the original parameter space or the transformed space, the variance of the trial distribution can be tweaked to adjust the average step size of the sampler. Nevertheless, it seems more intuitive to the writers to perform the sampling in the space in which adjusting the variance has an effect which is independent of where you are sampling from on the sphere.

A spherical distribution (namely the five-parameter Fisher-Bingham distribution, also known as the Kent distribution) is used in the toy model presented in Section~\ref{s:gns_tmII}.

\subsection{Variance of the Cartesian trial distribution}
\label{s:gns_spherevar}

For given variances of $\phi$ and $\theta$: $\sigma_{\phi}^2$ and $\sigma_{\theta}^2$, the variance corresponding to a function of these two variables is given by 
\begin{equation}
\label{e:gns_funcvar}
\sigma_{f}^2 = \left( \frac{\partial f}{\partial \phi} \right)^2 \sigma_{\phi}^2 + \left( \frac{\partial f}{\partial \theta} \right)^2 \sigma_{\theta}^2 + 2 \frac{\partial f}{\partial \phi} \frac{\partial f}{\partial \theta} \sigma_{\phi,\theta},
\end{equation}
where $\sigma_{\phi,\theta}$ is the covariance between $\phi$ and $\theta$. Hence one can calculate the corresponding variance in Cartesian coordinates, $\sigma_{x}^2$, $\sigma_{y}^2$, and $\sigma_{z}^2$ by substituting the equations given by~\ref{e:gns_spherecoords} into equation~\ref{e:gns_funcvar}. Using these values for $q(x',y',z'|x,y,z)$ however, leads to an asymmetric trial distribution in its arguments, since the variance is now a function of $\theta$ and $\phi$. Our entire formulation of the geometric nested sampling algorithm requires $q$ to be symmetric in order for equations~\ref{e:gns_ns_accept} and~\ref{e:gns_db} to hold. Thus we set $\sigma_{x}^2 = \sigma_{y}^2 = \sigma_{z}^2 = 4 / 100$ to ensure $q$ is symmetric.  

\subsection{Non-spherical coordinate transformations}
\label{s:gns_nonspheretrans}

The transformation of the trial sampling problem introduced in the previous Section need not be unique to the case of a sphere. Indeed, our implementation of geometric nested sampling includes the option to transform to Cartesian coordinates from circular or toroidal parameters. This is done in the same way as described for the spherical case, but with the relations given by equation~\ref{e:gns_spherecoords} replaced with the equivalent transformations for a circle or torus (see \citealt{kj_thesis}). \\
However, given the circular nature of the variables parameterising the points on a circle / torus, we do not think that performing coordinate transformations for these objects will give any advantages over using the wrapped trial distributions in the original parameter spaces. Hence in the applications considered in this paper, parameters which exhibit circular or toroidal properties will be sampled using the wrapped trial distribution, whilst those of a spherical nature will be sampled using the coordinate transformation methodology.

The coordinate transformation methodology can be applied to arbitrary geometries; however it is important to recognise that geometries which lack symmetry will in general be much more difficult to sample from without breaking the trial distribution symmetry requirement of the Metropolis acceptance ratio. This may lead to violation of detailed balance which is a \textit{sufficient} condition for a Markov chain to asymptotically converge to the target distribution (in this case the posterior). 

\section{Applications of geometric nested sampling}
\label{s:gns_applications}

We now apply the geometric nested sampling algorithm to models which include circular, toroidal and spherical parameters. We evaluate the algorithm's performance by plotting the posterior samples using \textsc{getdist}. 
For comparison we calculate posterior samples using \textsc{MultiNest} \citep{2009MNRAS.398.1601F}, a state of the art clustering nested sampling algorithm, effective on low dimensional problems. \\
We refer to the samples / distributions obtained from the geometric nested sampler as MG (Metropolis geometric nested sampling), 
and those obtained from \textsc{MultiNest} as MN. \\

\subsection{Toy model I: circular and toroidal distributions}
\label{s:gns_tmI}
The first toy models considered highlight the usefulness of the wrapping of the trial distribution used by the geometric nested sampler in the case of circular or toroidal parameters.
\subsubsection{Circular distribution}
\label{s:gns_tmI_c}
We first consider the problem of a one-dimensional circular distribution from which we would like to sample from. The model is parameterised by one variable $\phi$, which is defined on $[0, 2\pi]$. Referring back to Section~\ref{s:gns_priors} we take $\pi(\phi)$ to be uniform on $[0, 2\pi]$. For the likelihood function, we use the von Mises distribution introduced in Section~\ref{s:gns_circular} and defined by 
\begin{equation}
\label{e:gns_c}
\mathcal{L}_{c}\left(\phi | \mu, \sigma^2\right) = \frac{\exp(\cos(\phi - \pi - \mu)/ \sigma^2)}{2\pi I_{0}\left(\frac{1}{\sigma^2}\right)},
\end{equation}
where $\mu$ and $\sigma$ are the mean and standard deviation of the distribution, and $I_{0}(x)$ is the zeroth order modified Bessel function. Here we set $\mu = 0$ so that the peak of the posterior distribution is wrapped around $[0, 2\pi]$, and appears as two half peaks. We set the variance equal to $0.25$. \\
Since the problem involves the circular parameter $\phi$, the geometric nested sampling algorithm uses a wrapped trial distribution.

For this low-dimensional (one-dimensional) problem both MG and MN recover the correct distribution easily, even when the algorithms are run with a low number of livepoints ($ n_{l} = 50 $). We therefore consider a more complicated model in the form of a toroidal distribution in the next Section, to which the geometric argument of applying the wrapped trial distribution methodology stands equally well.

\subsubsection{Toroidal distribution}
\label{s:gns_tmI_t}

\begin{figure*}
  \begin{center}
  \includegraphics[ width=0.80\linewidth]{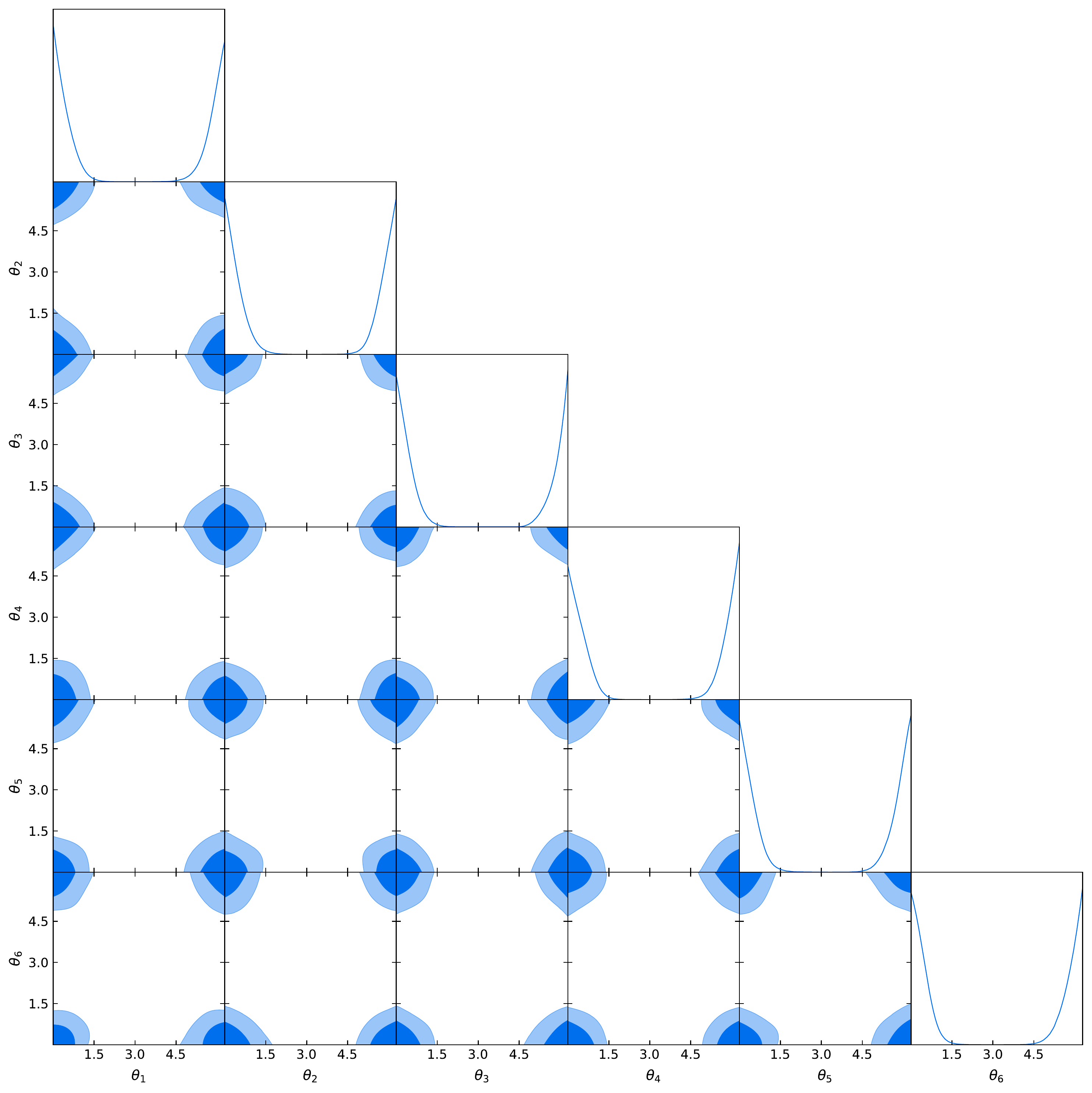}
  \caption{Posterior distribution of the 6-torus toy model defined in Section~\ref{s:gns_tmI_t} obtained with the geometric nested sampling algorithm, run with $50$ livepoints. The graphs along the upper diagonal are the one-dimensional marginalised posterior distributions for each of the 6-torus' parameters and show the 'half peaks' observed with circular distributions. The set of plots below the upper diagonal are the two-dimensional marginalised posterior distributions between pairs of $\theta_1,...,\theta_6$, and the light and dark blue shadings respectively correspond to the $95\%$ and $68\%$ mean confidence intervals. There are a total of $60$ `quarter peaks' across all pairs of parameters of the 6-torus.}\label{f:gns_l_t_post}
  \end{center}
\end{figure*}

The `usual' three-dimensional torus ($2$-torus) can be thought of as a topological space homeomorphic to the Cartesian product of two circles. The corresponding three-dimensional toroidal distribution is therefore the product of two circular distributions. This idea can be generalised to a hypertorus also known as an $n$-torus, in which case the $n$-toroidal distribution is equal to the product of $n$ circular distributions,
\begin{equation}
\label{e:gns_t}
\mathcal{L}_{t}\left(\theta_{1},...,\theta_{n} | \mu_{1},...,\mu_{n}, \sigma_{1}^2,...,\sigma_{n}^2\right) = \prod_{i=1}^{n}\mathcal{L}_{c}\left(\theta_{i} | \mu_{i}, \sigma_{i}^2\right).
\end{equation}
Here we use the same circular distribution as the one defined in Section~\ref{s:gns_tmI_c} and note that the `half peaks' observed in the one-dimensional parameter space of a circular distribution become 'quarter peaks' when observing any two-dimensional subspace of a toroidal parameter space. We thus expect this $n$-toroidal distribution to have $n (n-1) / 2 \times 4$ quarter peaks across all unique two-dimensional subspaces (see e.g. Figure~\ref{f:gns_l_t_post}) for $n\geq 2$, which most samplers would treat as a $2n (n-1)$ mode problem, while the geometric nested sampler would interpret them as just one mode. Note that when considering the entire parameter space at once, the toroidal distribution has $2^n$ modes and so for the 6-torus this corresponds to $64$ modes, but for purposes of visualisation we focus our analysis on the algorithms' ability to infer the two-dimensional quarter peaks.

\begin{figure*}
  \begin{center}
  \includegraphics[ width=0.80\linewidth]{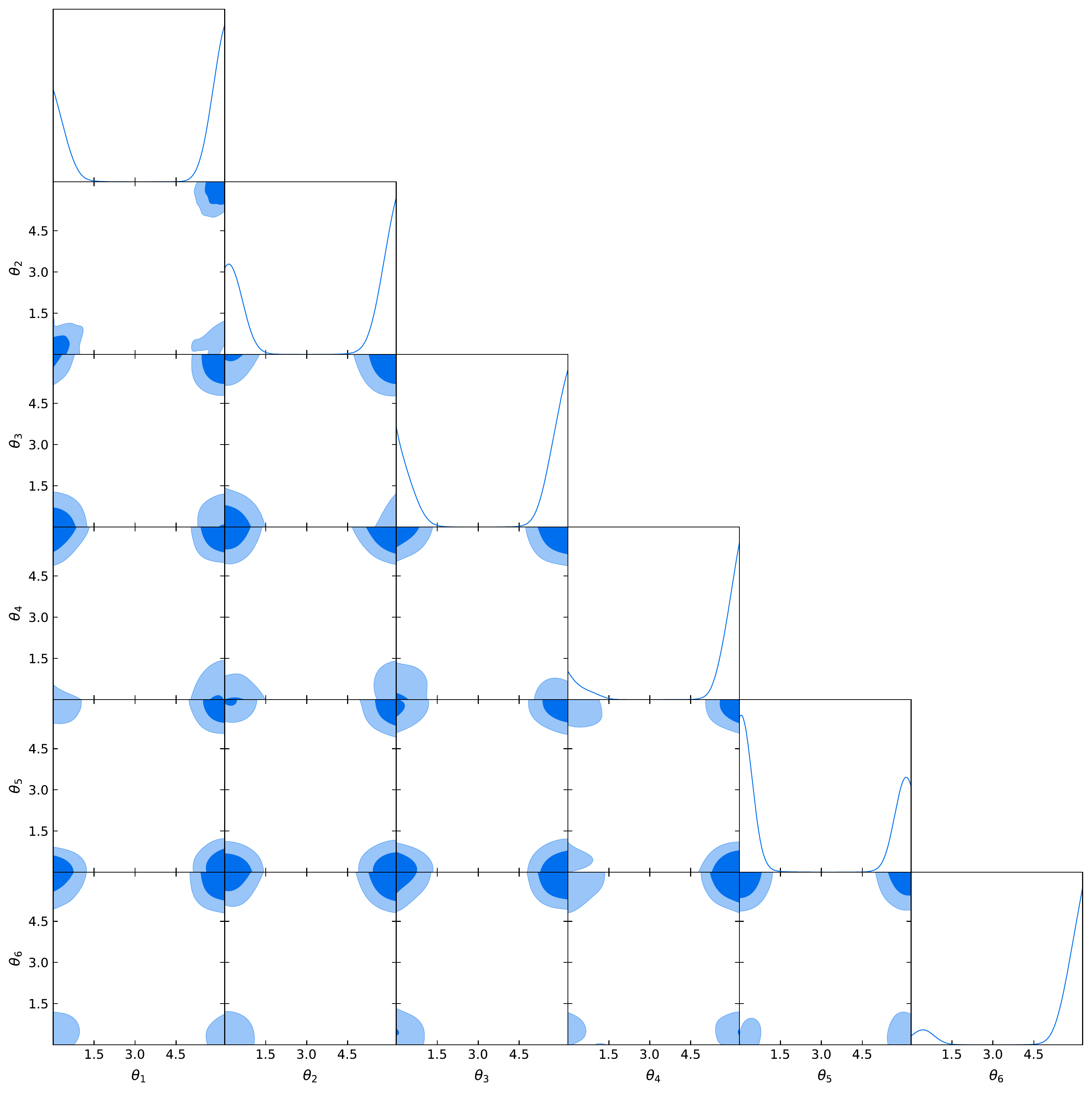}
  \caption{Posterior distribution of the 6-torus toy model defined in Section~\ref{s:gns_tmI_t} obtained with the MultiNest algorithm, run with $50$ livepoints. The layout is as described in Figure~\ref{f:gns_l_t_post}.}\label{f:mn_l_t_post}
  \end{center}
\end{figure*}

\begin{figure*}
  \begin{center}
  \includegraphics[ width=0.80\linewidth]{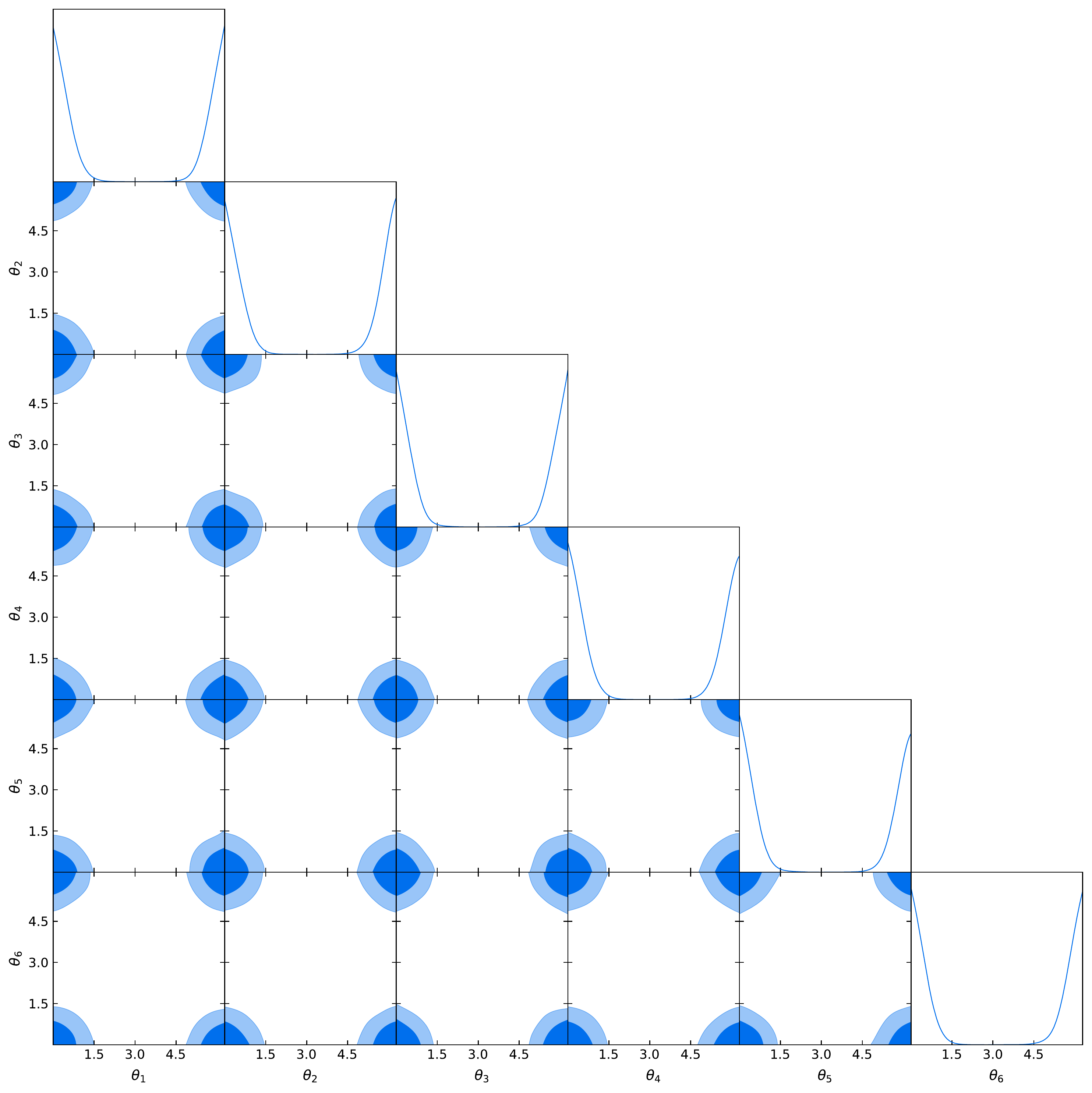}
  \caption{Posterior distribution of the 6-torus toy model defined in Section~\ref{s:gns_tmI_t} obtained with the geometric nested sampling algorithm, run with $500$ livepoints. The layout is as described in Figure~\ref{f:gns_l_t_post}.}\label{f:gns_h_t_post}
  \end{center}
\end{figure*}

\begin{figure*}
  \begin{center}
  \includegraphics[ width=0.80\linewidth]{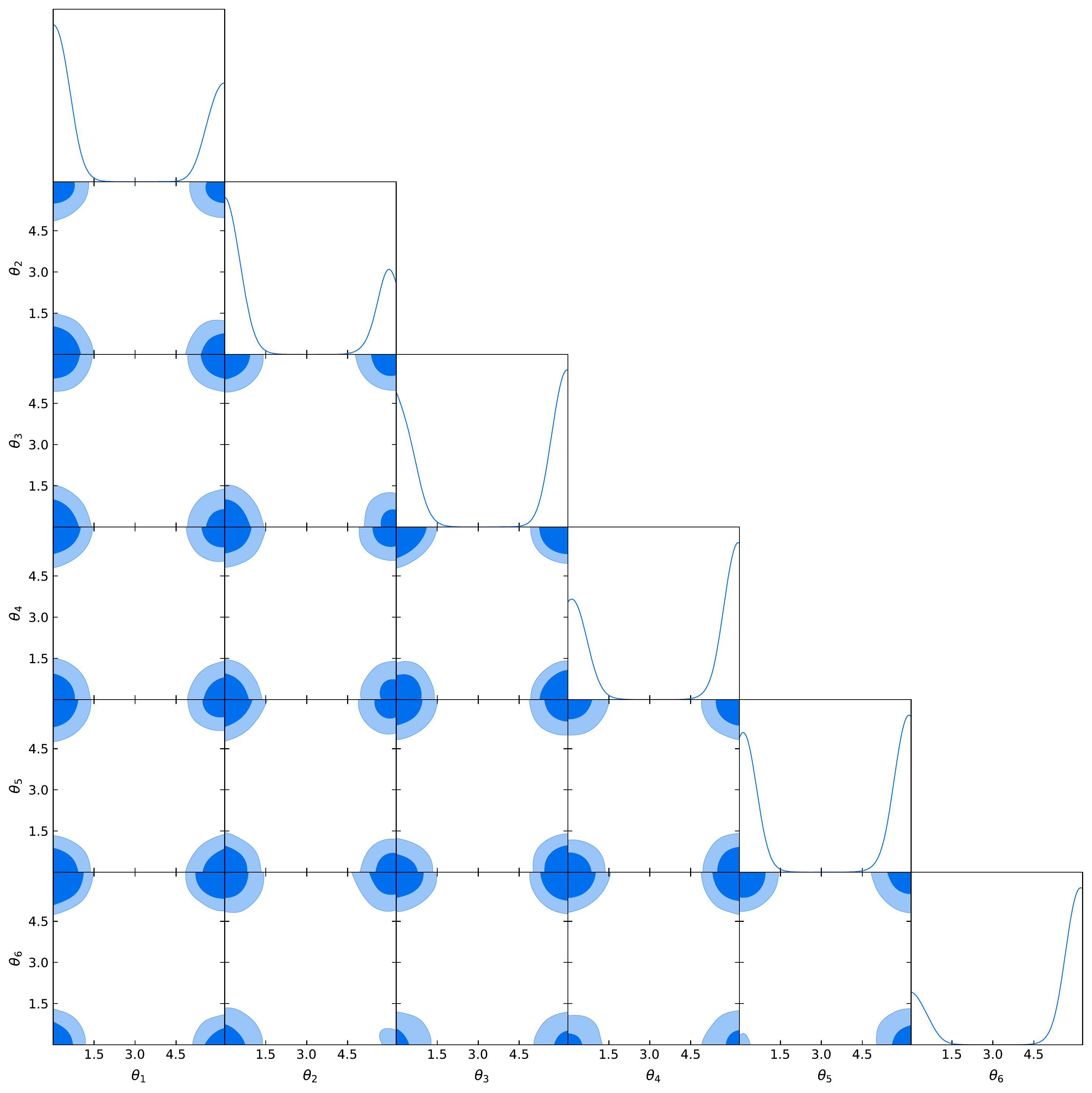}
  \caption{Posterior distribution of the 6-torus toy model defined in Section~\ref{s:gns_tmI_t} obtained with the MultiNest algorithm, run with $500$ livepoints. The layout is as described in Figure~\ref{f:gns_l_t_post}.}\label{f:mn_h_t_post}
  \end{center}
\end{figure*}

We run the MG and MN algorithms on a 6-torus toy model, which uses a uniform prior on $[0, 2\pi]$ for each $\theta_1,...,\theta_6$. For each sampler we do two runs, one with a low number of livepoints ($n_l = 50$) and one with a relatively high number ($n_l = 500$). \\
Figures~\ref{f:gns_l_t_post} and~\ref{f:mn_l_t_post} show the posterior distributions for the low livepoint runs using the MG and MN algorithms respectively. By looking at the two-dimensional marginalised posteriors it is clear that the MG sampler does a good job at recovering all $60$ quarter peaks present in the $6$-torus model. This is expected even for a low number of livepoints, as the MG still treats the problem as unimodal, and the dimensionality of the problem is not too taxing. On the other hand the MN algorithm appears to completely miss three quarter peaks in the $95\%$ mean confidence intervals (top left corner of $\theta_1-\theta_2$, bottom right corner of $\theta_2-\theta_6$, and bottom right corner of $\theta_3-\theta_6$ posteriors) while it significantly underestimates the posterior mass (does not capture the peaks in the $68\%$ contours) on a further $14$ of the quarter peaks. For a low number of livepoints this is to be expected for any sampler which treats each of the quarter peaks as separate modes. In fact MultiNest does a surprisingly good job at not completely missing peaks, given the low number of livepoints means it can't generate nearly as many ellipsoids as there are quarter peak modes, or more importantly the $64$ modes present when considering the parameter space as a whole. \\
Figures~\ref{f:gns_h_t_post} and~\ref{f:mn_h_t_post} show similar plots of the posteriors but with each algorithm run with $500$ livepoints. The MG algorithm shows little improvement over its performance with $50$ livepoints, which if anything emphasises how well the algorithm performed in the low livepoint case. The MultiNest algorithm shows a marked improvement over its performance using $50$ livepoints, as it no longer completely misses any quarter peaks in the $95\%$ contours, and only two peaks are not encapsulated by the $68\%$ contours. Nevertheless the algorithm still underperforms the $50$ livepoint run with the MG algorithm.

\subsection{Toy model II: spherical distribution}
\label{s:gns_tmII}

We now consider a spherical toy model to illustrate the Euclidean transformation technique described in Section~\ref{s:gns_coordtrans}. The geometric nested sampler uses this technique for variables which parameterise points on a sphere in terms of an azimuthal angle (usually denoted $\theta$) and polar angle ($\phi$). A natural choice for this model is the five parameter Fisher-Bingham distribution (\citealt{1982kent}, also known as the Kent distribution) which is described below.

\subsubsection{Kent distribution}
\label{s:gns_kent}

The Kent distribution is a probability distribution defined on the surface of a three-dimensional unit sphere. It is the spherical equivalent to a two-dimensional Gaussian distribution on a linear space, and can be parameterised in terms of spherical or Cartesian coordinates. In the latter case the distribution is given by
\begin{equation}
\label{e:gns_k}
\mathcal{L}_{K}\left(\hat{\vec{x}}\right) = \frac{1}{c(\kappa, \beta)} \exp \left( \kappa \hat{\vec{\gamma}}_1 \cdot \hat{\vec{x}} + \beta \left[ (\hat{\vec{\gamma}}_2 \cdot \hat{\vec{x}})^2 - (\hat{\vec{\gamma}}_3 \cdot \hat{\vec{x}})^2 \right] \right),
\end{equation}
(suppressing the parameters being conditioned on in $\mathcal{L}_{K}(\hat{\vec{x}})$ for brevity). Here $\hat{\vec{x}}$ is a unit vector pointing from the centre of the sphere to a point on its surface. The parameters $\kappa$ and $\beta$ describe the concentration (c.f. $1 / \sigma^2$ in the normal distribution) and ellipticity of the distribution respectively. The higher $\kappa$ ($\beta$) is, the more concentrated (elliptical) the contours of equal probability are. The vectors $\hat{\vec{\gamma}}_1,~\hat{\vec{\gamma}}_2$ and $\hat{\vec{\gamma}}_3$ describe the orientation of the distribution, with $\hat{\vec{\gamma}}_1$ pointing (from the centre of the sphere to a point on its surface) to the mean of the distribution, while $\hat{\vec{\gamma}}_2$ and $\hat{\vec{\gamma}}_3$ point in the direction of the major and minor axes of the contours of the distribution. Thus the three vectors must be orthogonal to each another. $c(\kappa, \beta)$ is a normalisation factor given by
\begin{equation}
\label{e:gns_k_const}
c(\kappa, \beta) = 2 \pi \sum_{i=0}^{i = \infty} \frac{\Gamma \left(i+ \frac{1}{2} \right)}{\Gamma \left(i + 1 \right)}\beta^{2i} \left( \frac{\kappa}{2} \right)^{-2i - \frac{1}{2}} I_{2i + \frac{1}{2}}(\kappa),
\end{equation}
where $I_{\alpha}(x)$ is the $\alpha^{\mathrm{th}}$ order modified Bessel function.

\begin{figure*}
  \begin{center}
  \includegraphics[ width=0.80\linewidth]{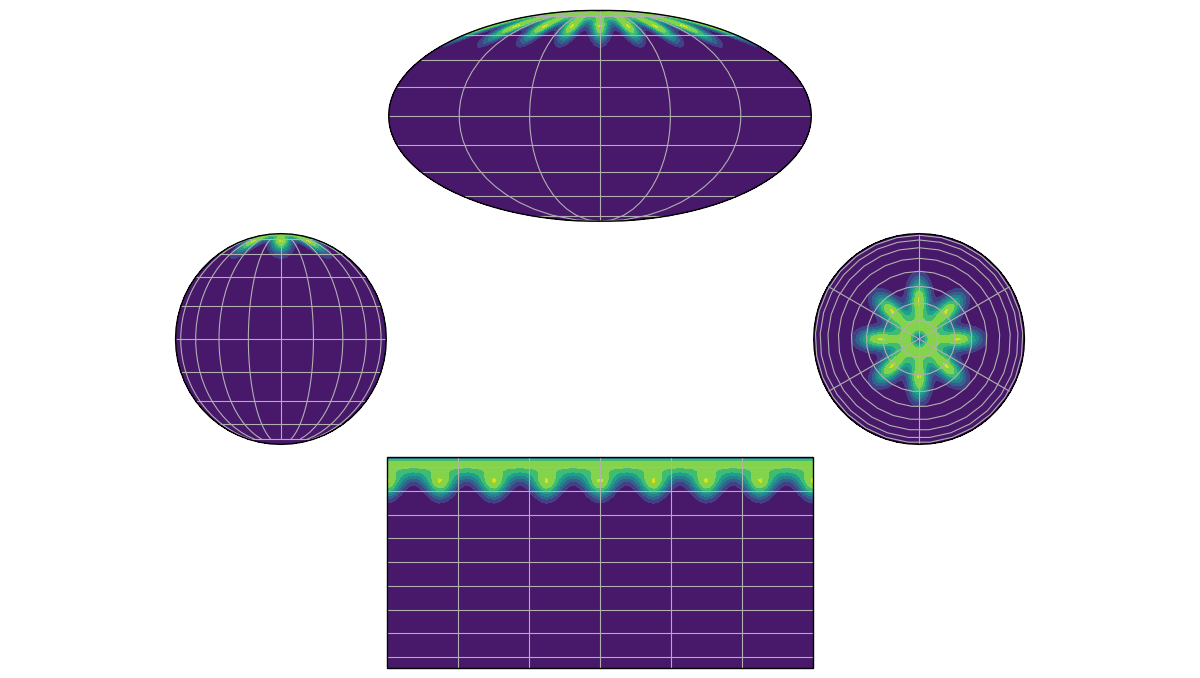}
  \caption{Posterior distribution of the spherical likelihood described by equations~\ref{e:gns_k},~\ref{e:gns_s} and~\ref{e:gns_gammas}, and a spherical prior (uniform in $\phi$ and sinusoidal in $\theta$), determined by evaluating the posterior over a grid of $\phi,~\theta$. The brighter regions correspond to regions of higher probability mass.}\label{f:gns_kentg100}
  \end{center}
\end{figure*}

The model we consider here is a sum of four Kent distributions, each with $\kappa = 100$ and $\beta = 50$. Defining $\mathbfss{G}_i$ as the matrix of orthogonal column vectors $\hat{\vec{\gamma}}_1,~\hat{\vec{\gamma}}_2$ and $\hat{\vec{\gamma}}_3$ for the $i^{\mathrm{th}}$ Kent distribution, the spherical likelihood is given by
\begin{equation}
\label{e:gns_s}
\mathcal{L}_{s}(\hat{\vec{x}}) = \sum_{i = 1}^{i = 4} \mathcal{L}_{K}\left(\hat{\vec{x}} | \mathbfss{G}_i \right),
\end{equation}
where we explicitly state the conditional dependence of $\mathcal{L}_{K}$ on $\mathbfss{G}_i$ to emphasise the matrices are different for each $\mathcal{L}_{K}$ in the summation. The $\mathbfss{G}_i$ are given by
\begin{equation}
\label{e:gns_gammas}
\begin{split}
\mathbfss{G}_1 
&= 
\begin{bmatrix}
0 & 0 & 1 \\
0 & 1 & 0 \\
1 & 0 & 0
\end{bmatrix},
\qquad 
\mathbfss{G}_2 
= 
\begin{bmatrix}
0 & 1 & 0 \\
0 & 0 & 1 \\
1 & 0 & 0
\end{bmatrix}, \\
\mathbfss{G}_3 
&= 
\frac{1}{\sqrt{2}}\begin{bmatrix}
0 & -1 & 1 \\
0 & 1 & 1 \\
\sqrt{2} & 0 & 0
\end{bmatrix},
\qquad 
\mathbfss{G}_4 
= 
\frac{1}{\sqrt{2}}\begin{bmatrix}
0 & 1 & -1 \\
0 & 1 & 1 \\
\sqrt{2} & 0 & 0
\end{bmatrix}.
\end{split}
\end{equation}
When combined with a uniform prior over $[0, 2\pi]$ for polar angle $\phi$ and sin prior over $[0, \pi]$ for azimuthal angle $\theta$, the resultant posterior is the one shown in Figure~\ref{f:gns_kentg100} (determined by evaluating the posterior over a grid of $\phi,~\theta$). We refer to this as the `flower' distribution, which is centred on `north' pole ($\theta = 0$) of the sphere. Note that the eight `petals' of the distribution more or less correspond to modes of the distribution when projected on a linear space (bottom plot of the Figure).
\subsubsection{Sampling the spherical toy model}
\label{s:gns_kent_results}
\begin{figure*}
  \begin{center}
  \includegraphics[ width=0.80\linewidth]{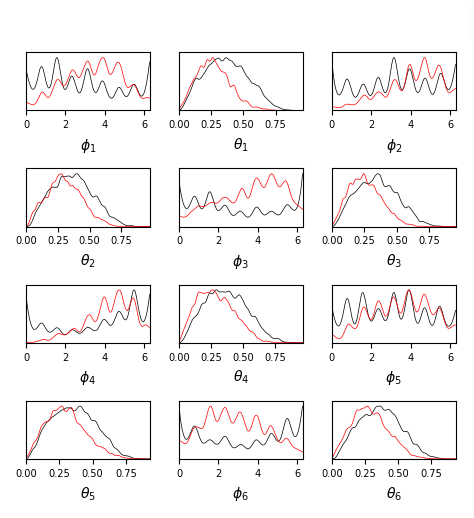}
  \caption{Posterior distribution of the spherical toy model consisting of six spheres each with a flower distribution and sinusoidal prior defined on their surfaces. Each plot shows the one-dimensional marginalised posteriors for each of the $12$ spherical parameters (two parameters per sphere). $\phi_i$ are the polar angles and $\theta_i$ are the azimuthal angles. The black curves are calculated from the geometric nested algorithm samples, while the red curves are calculated from samples obtained by MultiNest. Both algorithms were run with $500$ livepoints.}\label{f:gns_h_s_post}
  \end{center}
\end{figure*}

Due to the complexity of the flower distribution, we only consider runs of the MG and MN algorithms with $500$ livepoints.
We first ran the algorithms on a model involving a single sphere which translates to a two-dimensional sampling space in $\phi$ and $\theta$. However, as was the case with the two-dimensional circular distribution, both algorithms recovered the posterior distribution with ease due to the low dimensionality of the problem. We therefore consider the problem of six separate spheres (in the case of the geometric nested sampler, the angles parameterising each sphere are transformed individually) each containing a petal distribution as the likelihood and a spherical prior. This results in a $12$-dimensional parameter space: $\phi_1, \theta_1,..., \phi_6, \theta_6$ which should prove significantly more challenging for the samplers. \\
Figure~\ref{f:gns_h_s_post} shows the results for the MG (black curves) and MN (red curves). Neither algorithm by any means recovers the true distribution perfectly, but for the majority of the polar angles the MG infers the eight petals of the distribution more symmetrically than MN, with the exception of $\phi_1$ and $\phi_6$ where both algorithms perform poorly. This toy model presents a problem with $6 \times 8 = 48$ modes, a modest amount, but it is the geometric nature of these modes which causes difficulties for MN. Referring back to Figure~\ref{f:gns_kentg100} encapsulating the areas of prior mass is difficult for ellipsoidal samplers due to the petals being connected to each other, as it requires lots of small ellipsoids to be concatenated together to approximate the shape of the region. Combining this with the higher dimensionality of the problem of six spheres means that regions of high probability are inevitably going to be missed by the algorithm. The fact that the distribution is centred around the pole of the sphere hinders algorithms which sample in the angular parameter space further, as such an algorithm must move `around' the pole to sample the different petals. The geometric nested sampler does not suffer from this issue, as since it samples from the Cartesian space, it has no concept of the pole of the sphere.

\section{Geometric nested sampling implementation}
\label{s:gns_imp}
The implementation of the geometric nested sampler (and the vanilla Metropolis nested sampler) used in this paper, along with the toy models and the gravitational wave likelihood function can be found at \url{https://github.com/SuperKam91/nested_sampling}. The algorithm is written in \textsc{Python 2.7}, hence our \textit{implementation} of the algorithm cannot match that of the state of the art nested sampling algorithms such as \textsc{MultiNest} or \textsc{POLYCHORD} \citep{2015MNRAS.453.4384H}. These algorithms are implemented in \textsc{FORTRAN 90}, and parallelised using a master-slave paradigm (see Section~5.4 of Handley, Hobson, \& Lasenby). Nevertheless there is no reason why geometric nested sampling cannot be implemented more efficiently and parallelised using this method. 

\section{Conclusions}
\label{s:gns_conc}

We have presented a new nested sampling algorithm based on the Metropolis nested sampler proposed in \citet{Sivia2006} and applied in \citet{2008MNRAS.384..449F}. Our algorithm exploits the geometric properties of certain kinds of parameters which describe points on circles, tori and spheres, to sample the parameters more efficiently in the context of nested sampling. The algorithm should be more mobile in sampling distributions defined on such geometries. \\
The algorithm consists of two key sampling modes which can be summarised as follows.
\begin{itemize}
\item For circular and toroidal problems, the trial distribution used in the sampling process is wrapped around the support of the prior distribution $\pi$ (domain of the posterior distribution $\mathcal{P}$).
\item This wrapping ensures that no trial points are automatically rejected when evaluating the Metropolis acceptance ratio as a consequence of the point being outside the sampling space of the model.
\item The wrapped trial distribution also makes the sampling more mobile at the edges of the domain of $\mathcal{P}$, meaning that circular and toroidal distributions should be easier to sample, particularly in the case of posteriors with high probability densities at these edges.
\item For spherical problems, parameters specifying the coordinates on a sphere are transformed to Cartesian coordinates and sampled from the corresponding Euclidean space.
\item This again ensures that no trial points are automatically rejected because they are outside the domain of $\mathcal{P}$.
\item It also enhances the mobility of the sampler, whose average step size along the surface of the sphere is not dependent on the location at which the trial distribution is centred. 
\end{itemize}
We applied the geometric nested sampling algorithm (MG) to three toy models, which respectively represented models on a circle, hypertorus ($n$-torus) and spheres. We compared the posterior plots with those obtained from 
the livepoint clustering nested sampling algorithm \textsc{MultiNest} (MN, \citealt{2009MNRAS.398.1601F}). 
Our results can be summarised as follows.
\begin{itemize}
\item For the circular toy model (von Mises distribution centred on the origin), the MG and MN samplers perform equally well. We attribute this to the low dimensionality of the problem.
\item We therefore considered a toroidal ($6$-torus) distribution which was equivalent to the product of six von Mises circular distributions. 
\item We found that when using a low number of livepoints ($50$), the MG recovers all $60$ quarter peaks present in two-dimensional parameter subspaces very well, while the MN algorithm more or less completely missed three of these peaks, and recovered a further $14$ of them poorly. 
\item For a high livepoint run ($500$) the MG performs similar as to what it did in the low livepoint case, while the MN shows significant improvement but still underperforms relative to the low livepoint MG run. This is in essence because the MG treats the problem as a unimodal distribution.
\item The first spherical toy model considered comprised of a sum of four Kent distributions (a normal distribution defined on the surface of a unit $2$-sphere) whose resultant distribution resembles a `flower' centred on the pole of the sphere, which had eight `petals' containing high probability mass. This two-dimensional problem in polar angle $\phi$ and azimuthal angle $\theta$ proved easy for both samplers, due to low dimensionality.
\item We therefore considered a problem comprised of six separate spheres, each of which had its own flower distribution and was transformed to Euclidean space independently in the case of the geometric nested sampler. This translates to a $12$-dimensional parameter estimation problem in $\phi_1, \theta_1,..., \phi_6, \theta_6$ with $6 \times 8 = 48$ petals (modes of the distribution).
\item We found that both algorithms failed to recover the true distributions near perfectly, but the MG did a better job at inferring the correct shape of the modes (and their symmetry with respect to each other).
\end{itemize}

\section*{Acknowledgements}
The authors would like to thank 
Greg Willatt \& David Titterington from Cavendish Astrophysics for computing assistance. They would also like to thank Dave Green for his invaluable help using \LaTeX.
Kamran Javid acknowledges an STFC studentship. 

\setlength{\bibsep}{0pt}            
\renewcommand{\bibname}{References} 


\bsp	
\label{lastpage}
\end{document}